\documentclass{article}
\usepackage{spconf,amsmath,graphicx}
\usepackage{multirow}
\usepackage{url}
\usepackage{cite}
\usepackage{subcaption}
\usepackage{cool}
\usepackage{booktabs} 

\title{Speech and Speaker Recognition from Raw Waveform with SincNet}
\name{Mirco Ravanelli, Yoshua Bengio$^{*}$}
\address{
  Mila, Universit\'e de Montr\'eal , $^*$CIFAR Fellow}

\begin{document}
\ninept
\maketitle
\begin{abstract}
Deep neural networks can learn complex and abstract representations, that are progressively obtained by combining simpler ones.  A recent trend in speech and speaker recognition consists in discovering these representations starting from raw audio samples directly.  Differently from standard hand-crafted features such as MFCCs or FBANK, the raw waveform can potentially help neural networks discover better and more customized representations. The high-dimensional raw inputs, however, can make training significantly more challenging. 

This paper summarizes our recent efforts to develop a neural architecture that efficiently processes speech from audio waveforms. In particular, we propose \textit{SincNet}, a novel Convolutional Neural Network (CNN) that encourages the first layer to discover meaningful filters by exploiting parametrized sinc functions. In contrast to standard CNNs, which learn all the elements of each filter, only low and high cutoff frequencies of band-pass filters are directly learned from data. This inductive bias offers a very compact way to derive a customized front-end, that only depends on some parameters with a clear physical meaning. 

Our experiments, conducted on both speaker and speech recognition, show that the proposed architecture converges faster, performs better, and is more computationally efficient than standard CNNs.
\end{abstract}

\section{Introduction}
\label{sec:intro}
Deep learning has shown remarkable success in numerous speech tasks \cite{Goodfellow-et-al-2016-Book}, including speech \cite{lideng,ravanelli_thesis} and speaker recognition \cite{dnn_speaker_rec2}. 
This paradigm exploits the principle of compositionality to efficiently describe the world around us and employs a hierarchy of representations that are progressively learned by combining lower-level abstractions. To fully take advantage of deep learning, it would thus be natural to directly feed neural networks with the lowest possible signal representation (e.g., pixel for images or raw samples for audio), avoiding any kind of pre-computed intermediate representations.

Nevertheless, most state-of-the-art neural networks used for speech applications still employ hand-crafted features, such as FBANK and MFCC coefficients. These engineered features are originally designed from perceptual evidence and there are no guarantees that such representations are optimal for all speech-related tasks. Standard features, for instance, smooth the speech spectrum, possibly hindering the extraction of crucial narrow-band speaker characteristics such as pitch and formants. To mitigate this drawback, some recent works have proposed to directly feed the network with spectrogram bins \cite{e2e_spk_id,spk_rec_time_freq,voxceleb} or even with raw waveforms \cite{palaz_raw,tara_raw,tuske,dnn_emotion,acoustic_raw_povey,spoofing_raw,raw_speaker_id,verification_raw_ICASSP2018,verification_raw_IS2018}. CNNs are the  best candidate for processing raw speech samples, since weight sharing, local filters, and pooling help discover robust and invariant representations. 

We believe that one of the most critical parts of current waveform-based CNNs is the first convolutional layer. This layer not only deals with high-dimensional inputs, but it is also more affected by vanishing gradient problems, especially when employing very deep architectures. As we will show in this paper, the filters learned by CNNs often take noisy and incongruous multi-band shapes, especially when few training samples are available. These filters certainly make some sense for the neural network, but do not appeal to human intuition, nor appear to lead to an efficient representation of the speech signal. 

To help the CNNs discover more meaningful filters, we recently proposed a novel convolutional architecture, called SincNet \cite{SincNet}, that adds some constraints on the filter shape. Compared to standard CNNs, where the filter-bank characteristics depend on several parameters (each element of the filter vector is directly learned), SincNet convolves the waveform with a set of parametrized sinc functions that implement band-pass filters \cite{SincNet}. The low and high cutoff frequencies of the filters are the only parameters learned from data. This solution still offers considerable flexibility but forces the network to focus on high-level tunable parameters that have a clear physical meaning. 

In \cite{SincNet} we obtained promising results on both speaker identification and speaker verification tasks. In particular, SincNet turned out to outperform standard CNNs fed by both raw samples and standard features. Motivated by these encouraging results, this paper extends our recent findings to speech recognition tasks. 
The speech recognition experiments conducted here are based on the  TIMIT and DIRHA \cite{dirha_asru,rav_is16} datasets.  
Results confirm that the proposed SincNet converges faster, achieves better performance, is more computationally efficient, and leads to more interpretable filters than a standard CNN. Interestingly, SincNet achieves a competitive performance also on the DIRHA dataset, highlighting the effectiveness of this approach even in challenging scenarios characterized by the presence of both noise and reverberation. 

\section{The SincNet Architecture} \label{sec:sinc}
The first layer of a standard CNN performs a set of time-domain convolutions between the input waveform and some Finite Impulse Response (FIR) filters \cite{rabiner11}. Each convolution is defined as follows:
\begin{equation}
y[n]=x[n]*h[n] = \sum\limits_{l=0}^{L-1} x[l]\cdot h[n-l],
\end{equation}
where $x[n]$ is a chunk of the speech signal, $h[n]$ is the filter of length $L$, and $y[n]$ is the filtered output. In standard CNNs, all the L elements (taps) of each filter are learned from data. Conversely, the proposed SincNet performs the convolution with a predefined function $g$ that depends on few learnable parameters $\theta$ only, as highlighted in the following equation:

\begin{equation}
y[n]=x[n]*g[n,\theta]. 
\end{equation}

A reasonable choice, inspired by standard filtering in digital signal processing, is to define $g$ in order to employ rectangular bandpass filters. In the frequency domain, the magnitude of a generic bandpass filter can be written as the difference between two low-pass filters:

\begin{equation}
G[f,f_1,f_2]= rect\Big(\frac{f}{2f_{2}}\Big) - rect\Big(\frac{f}{2f_{1}}\Big),
\end{equation}
where $f_{1}$ and $f_{2}$ are the learned low and high cutoff frequencies, and $rect(\cdot)$ is the rectangular function in the magnitude frequency domain\footnote{The phase of the $rect(\cdot)$ function is considered to be linear.}.
After returning to the time domain (using the inverse Fourier transform \cite{rabiner11}), the reference function $g$ becomes:

\begin{equation}
g[n,f_1,f_2]= 2f_{2}sinc(2\pi f_2n) - 2f_{1}sinc(2\pi f_1n),
\end{equation}
where the sinc function is defined as $sinc(x)=sin(x)/x$. 

The cut-off frequencies can be initialized randomly in the range $[0,f_s/2]$, where $f_s$ represents the sampling frequency of the input signal.  
As an alternative, filters can be initialized with the cutoff frequencies of the mel-scale filter-bank, which has the advantage  of directly allocating more filters in the lower part of the spectrum, where many crucial speech information is located. Note that the gain of each filter is not learned at this level. This parameter is managed by the subsequent layers, which can easily attribute more or less importance to each filter output. 





An ideal bandpass filter (i.e., a filter where the passband is perfectly flat and the attenuation in the stopband is infinite) requires an infinite number of elements $L$. Any truncation of $g$ thus inevitably leads to an approximation of the ideal filter, characterized by ripples in the passband and limited  attenuation  in  the  stopband.   A popular solution to mitigate this issue is windowing \cite{rabiner11}. Windowing is performed by multiplying the truncated function $g$ with a window function $w$, which aims to smooth out the abrupt discontinuities at the  ends  of  $g$.
This paper uses the popular Hamming window that is particularly suitable to achieve high frequency selectivity \cite{rabiner11}. However, results not reported here reveals no significant performance difference when adopting other functions, such as Hann, Blackman, and Kaiser windows.

All operations involved in SincNet are fully differentiable and the cutoff frequencies of the filters can be jointly optimized with other CNN parameters using Stochastic Gradient Descent (SGD) or other gradient-based optimization routines. 
A standard CNN pipeline (pooling, normalization, activations, dropout) can be employed after the first sinc-based convolution.
Multiple standard convolutional, fully-connected or recurrent layers \cite{ravanelli_is17,li_gru,ravanelli_twin} can then be stacked together to finally perform a  classification with a softmax classifier. 

\begin{figure}[t!]
\begin{subfigure}{0.248\textwidth}
\includegraphics[scale=0.37,trim={1.8cm 0cm 0cm 0cm},clip]{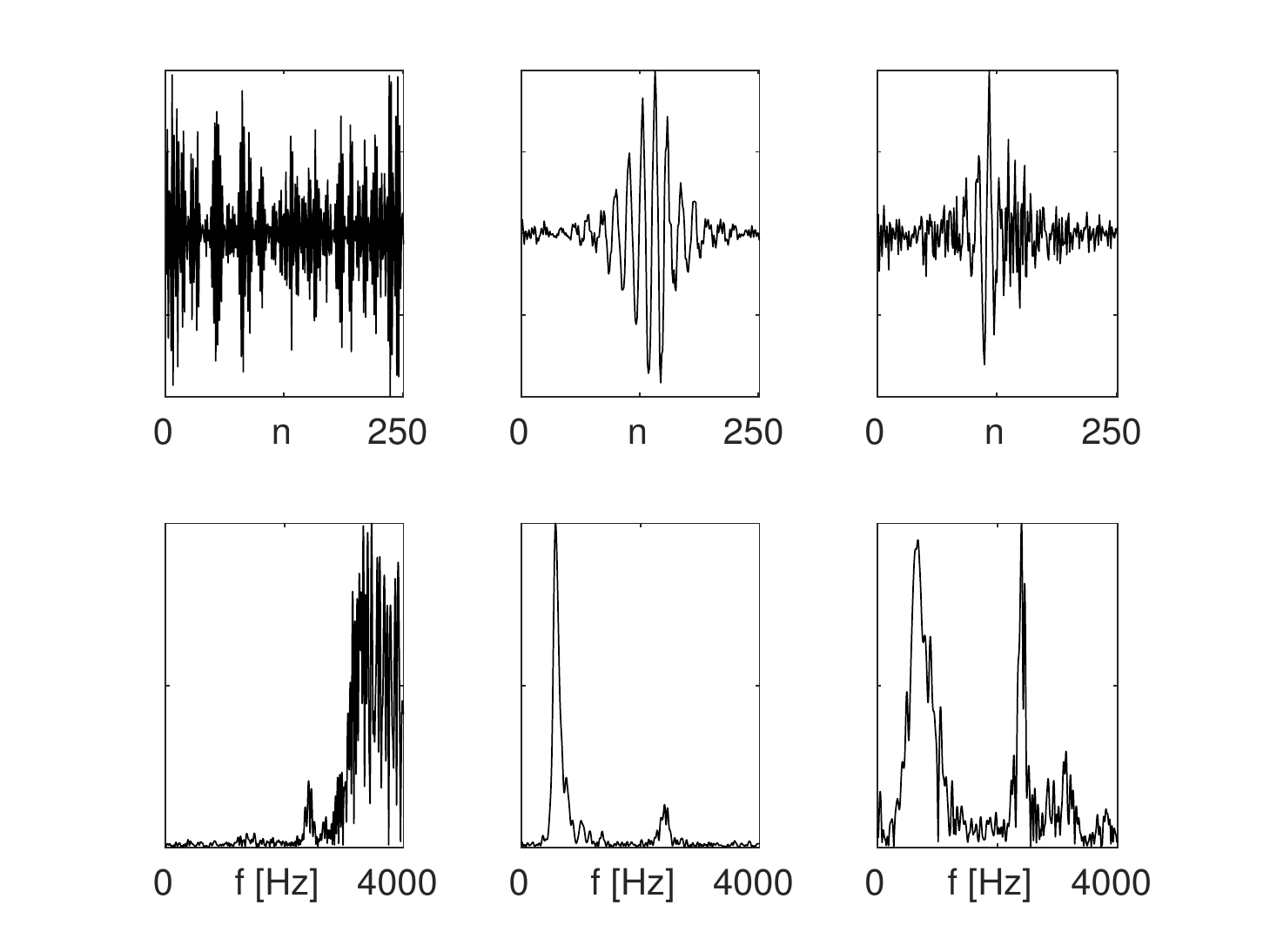}
\caption{CNN Filters}
\label{fig:cnn_filt}
\end{subfigure} \hspace{0.0\textwidth}
\begin{subfigure}{0.248\textwidth}
\includegraphics[scale=0.37,trim={1.8cm 0cm 0cm 0cm},clip]{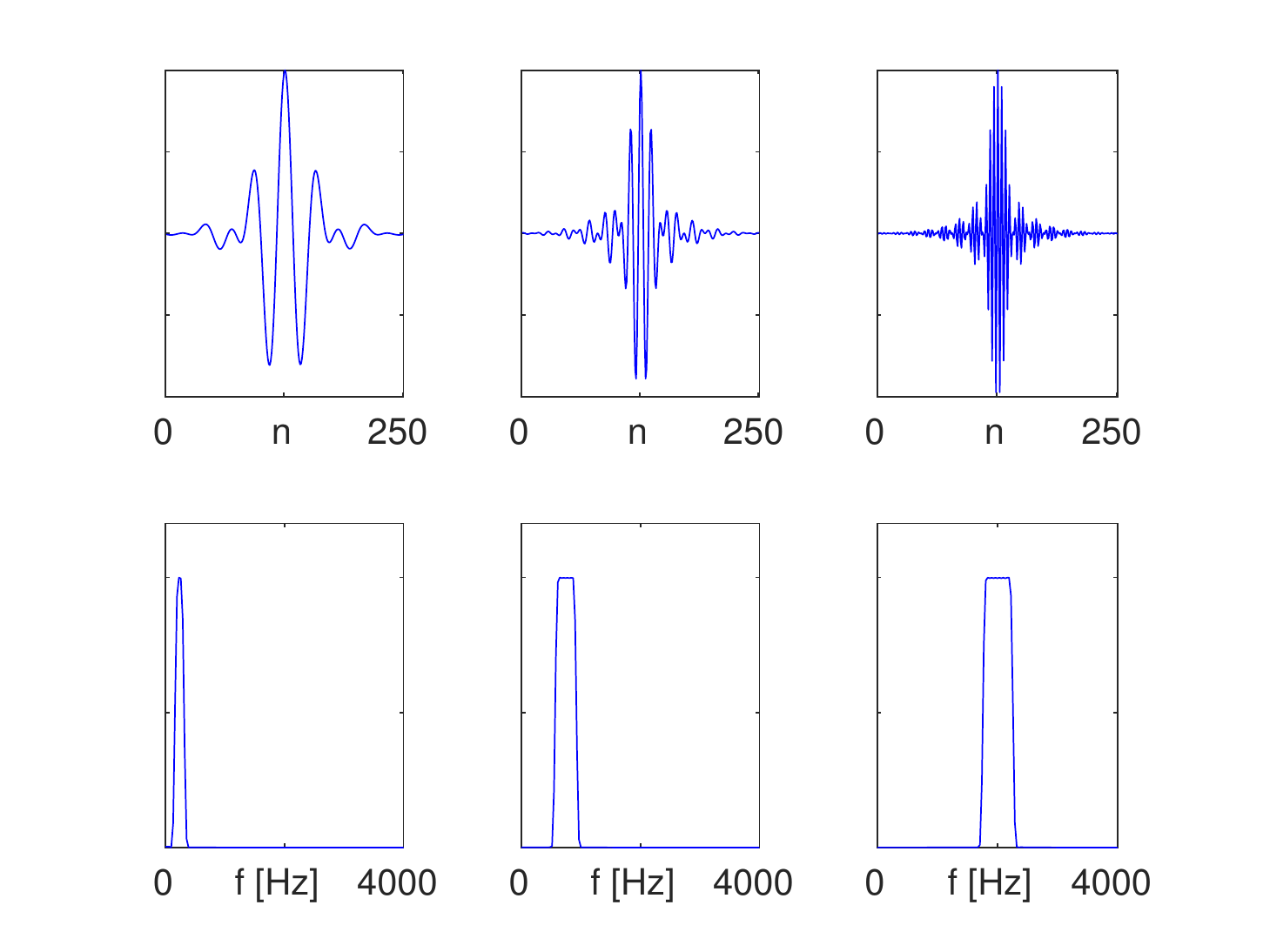}
\caption{SincNet Filters}
\label{fig:sinc_filt}
\end{subfigure}
\caption{Filters learned by a standard CNN and by the proposed SincNet. The first row reports the filters in the time domain, while the second one shows their magnitude frequency response.}
\label{fig:ir}
\end{figure}

Fig. \ref{fig:ir} shows some examples of filters learned by a standard CNN and by the proposed SincNet for a speaker identification task trained on Librispeech (the frequency response is plotted between 0 and 4 kHz). As observed in the figures, the standard CNN does not always learn filters with a well-defined frequency response. In some cases, the frequency response looks noisy (see the first CNN filter), while in others it assumes  multi-band shapes (see the third CNN filter). SincNet, instead, is specifically designed to implement rectangular bandpass filters, leading to more meaningful CNN filters. 

\subsection{Model properties}

The proposed SincNet has some remarkable properties:
\begin{itemize}
\item \textbf{Fast Convergence:}
SincNet forces the network to focus only on the filter parameters with a major impact on performance. Our architecture actually implements a natural inductive bias, utilizing knowledge about the filter shape (similar to feature extraction methods generally deployed on this task) while retaining flexibility to adapt to data. This prior knowledge makes learning the filter characteristics much easier, helping SincNet to converge significantly faster to a better solution.

\item \textbf{Few Parameters:} SincNet drastically reduces the number of parameters in the first convolutional layer. 
For instance, if we consider a layer composed of $F$ filters of length $L$, a standard CNN employs $F \cdot L$ parameters, against the $2F$ considered by SincNet. If $F=80$ and $L=100$, we employ 8k parameters for the CNN and only 160 for SincNet. Moreover, if we double the filter length $L$, a standard CNN doubles its parameter count (e.g., we go from 8k to 16k), while SincNet has an unchanged parameter count. This allows one to derive very selective filters with many taps, without adding learnable parameters. 

\item \textbf{Computational Efficiency}: The proposed function $g$ is symmetric. This means that we can perform convolution in a very efficient way by only considering one side of the filter and inheriting the results for the other half. This saves $50\%$ of the first-layer computation over a standard CNN. 

\item \textbf{Interpretability}: The SincNet feature maps obtained in the first convolutional layer are definitely more interpretable and human-readable than other approaches. The filters, in fact, only depend on parameters with a clear physical meaning.

\end{itemize}

\begin{figure}[t!]
\centering
  \includegraphics[scale=.50]{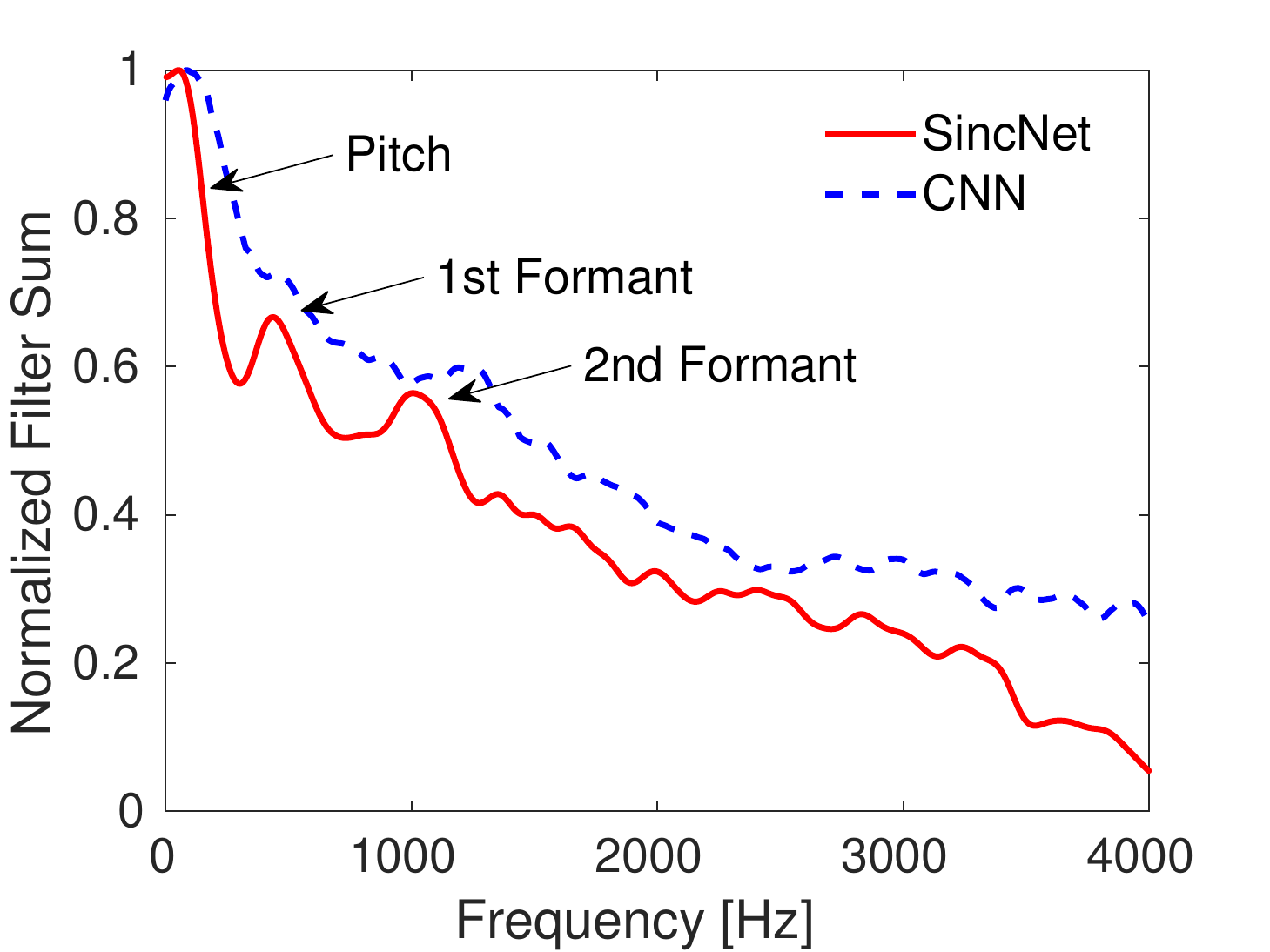}
  \caption{Cumulative frequency response of SincNet and CNN filters on speaker-id.}
  \label{fig:cum}
\end{figure}

\section{Related Work} \label{sec:rel_work}
Several works have recently explored the use of low-level speech representations to process audio and speech with CNNs. Most prior attempts exploit magnitude spectrogram features \cite{e2e_spk_id,spk_rec_time_freq,voxceleb,tara_asru2013,learn_fbank_const,fbank_par}. Although spectrograms retain more information than standard hand-crafted features, their design still requires careful tuning of some crucial hyper-parameters, such as the duration, overlap, and typology of the frame window, as well as the number of frequency bins. For this reason, a more recent trend is to directly learn from raw waveforms, thus completely avoiding any feature extraction step. 
This approach has shown promising in speech \cite{palaz_raw,tara_raw,google_rawmulti,joint7,tuske}, including emotion tasks \cite{dnn_emotion}, speaker recognition \cite{raw_speaker_id}, and spoofing detection \cite{spoofing_raw}.
Similar to SincNet, some previous works have proposed to add constraints on the CNN filters, for instance forcing them to work on specific bands \cite{tara_asru2013,learn_fbank_const}.  

Differently from the proposed approach, the latter works operate on spectrogram features and still learn all the L elements of the CNN filters. An idea related to the proposed method has been recently explored in \cite{fbank_par}, where a set of parameterized Gaussian filters are employed. This approach operates on the spectrogram domain, while SincNet directly considers the raw time domain waveform. Similarly to our work, in \cite{Neil_ICASSP2018} the convolutional filters are initialized with a predefined filter shape.  However, rather than focusing on cut-off frequencies only,  all the basic taps of the FIR filters are still learned.
 
This work extends our previous studies on the SincNet \cite{SincNet}. To the best of our knowledge,  this paper is the first that shows the effectiveness of this architecture in a speech recognition application. 

\section{Results} \label{sec:exp}
This section first summarizes the adopted experimental setup. We then discuss our recent speaker recognition results achieved on TIMIT (462 spks)  and Librispeech (2484 speakers) \cite{librispeech} datasets and then we extend the analysis to ASR (for more details on the speaker recognition experiments, see \cite{SincNet}).
Beyond adopting TIMIT, speech recognition experiments also consider the DIRHA dataset for assessing SincNet in challenging noisy and reverberant conditions \cite{dirha_asru}.
In the spirit of reproducible research, we release the code of SincNet on GitHub \cite{pytorch_kaldi}\footnote{\label{foot:code} at \url{https://github.com/mravanelli/SincNet/}.}. 

\subsection{SincNet Setup}
The waveform of each speech sentence was split into chunks of 200 ms (with 10 ms overlap), which were fed into the SincNet architecture. The first layer performs sinc-based convolutions as described in Sec. \ref{sec:sinc}, using 80 filters of length $L=251$ samples. The architecture then employs two standard convolutional layers, both using 60 filters of length 5. Layer normalization \cite{layer_norm} was used for both the input samples and for all convolutional layers (including the SincNet input layer). Next, three fully-connected layers composed of 2048 neurons and normalized with batch normalization \cite{batchnorm,ravanelli_SLT} were applied. All hidden layers used leaky-ReLU non-linearities. The parameters of the sinc-layer were initialized using mel-scale cutoff frequencies, while the rest of the network was initialized with the well-known ``Glorot" initialization scheme \cite{xavier}. Frame-level speaker and phoneme classifications were obtained by applying a softmax classifier, providing a set of posterior probabilities over the targets. For speaker-id, a sentence-level classification was simply derived by averaging the frame predictions and voting for the speaker which maximizes the average posterior. Training used the RMSprop optimizer, with a learning rate $lr=0.001$, $\alpha=0.95$, $\epsilon=10^-7$, and minibatches of size 128. 
All the hyper-parameters of the architecture were tuned on development data.
Please, refer to the GitHub repository for more details on the SincNet setup.



We compared SincNet with several alternative systems. 
First, we considered a standard CNN fed by the raw waveform. This network is based on the same architecture as SincNet, but replacing the sinc-based convolution with a standard one. 
A comparison with popular hand-crafted features was also performed. To this end, we computed 39 MFCCs (13 static+$\Delta$+$\Delta\Delta$) and 40 FBANKs using the Kaldi toolkit \cite{kaldi_short}. A CNN was used for FBANK features, while a Multi-Layer Perceptron (MLP) was used for MFCCs. 

\subsection{Speaker Recognition}
Fig. \ref{fig:cum} shows the cumulative frequency response of the filters learned by SincNet and CNN on a speaker-id task trained with Librispeech. The cumulative frequency response is obtained by summing up all the discovered filters and is useful to highlight which frequency bands are covered by the learned filters.

Interestingly, there are three main peaks which clearly stand out from the SincNet plot (see the red line in the figure). The first one corresponds to the pitch region (the average pitch is 133 Hz for a male and 234 for a female). The second peak (approximately located at 500 Hz) mainly captures first formants, whose average value over the various English vowels is indeed 500 Hz. Finally, the third peak (ranging from 900 to 1400 Hz) captures some important second formants, such as the second formant of the vowel $/a/$, which is located on average at 1100 Hz.
This filter-bank configuration indicates that SincNet has successfully adapted its characteristics to speaker identification. Conversely, the CNN does not exhibit such a meaningful pattern: its filters tend to focus on the lower part of the spectrum, but peaks tuned on first and second formants do not clearly appear. 

\begin{table}[h]
\caption{Performance of speaker identification (first two columns) and speaker verification (last column) on the considered corpora.}
\centering

\begin{tabular}{lllr}  
\toprule
& TIMIT & LibriSpeech & LibriSpeech  \\ 
& CER(\%) & CER(\%) & EER(\%)  \\ 
\midrule
DNN-MFCC             &   0.99      &  2.02   & 0.88   \\ 
CNN-FBANK       &   0.86      &  1.55  & 0.60    \\ 
CNN-Raw       &   1.65     &  1.00   & 0.58    \\ 
SincNet-Raw       &   \textbf{0.85}      &  \textbf{0.96}  &  \textbf{0.51}     \\ 
\bottomrule
\end{tabular}
\label{tab:spk_id_res}
\end{table}

Table \ref{tab:spk_id_res} reports the performance achieved on speaker identification and verification tasks (Classification Error Rates - CER\% for speaker-id task and Equal Error Rate - EER\% for speaker verification). The table shows that  SincNet outperforms other systems on both TIMIT (462 speakers) and Librispeech (2484 speakers) datasets. 
The speaker verification system was derived from the speaker-id neural network using the \textit{d-vector} approach \cite{dnn_spk_rec_class2,voxceleb}, which relies on the output of the last hidden layer and computes the cosine  distance between the enrollment and the test speaker d-vectors (see \cite{SincNet} for more details).
Ten utterances from impostors were randomly selected for each sentence from a genuine speaker. 
Note that to assess our approach on a standard open-set speaker verification task, all the impostors were taken from a speaker pool different from that used for training the speaker-id DNN.
The last column of Table \ref{tab:spk_id_res} extends our validation to speaker verification, reporting the EER(\%) achieved with Librispeech. All DNN models show promising performance, leading to an EER lower than 1\% in all cases. The table highlights that SincNet outperforms the other models, showing a relative improvement of about 11\% over the standard CNN.

\begin{figure}[t!]

\begin{subfigure}{0.2\textwidth}
\includegraphics[scale=0.555,trim={0cm 0cm 0cm 0cm},clip]{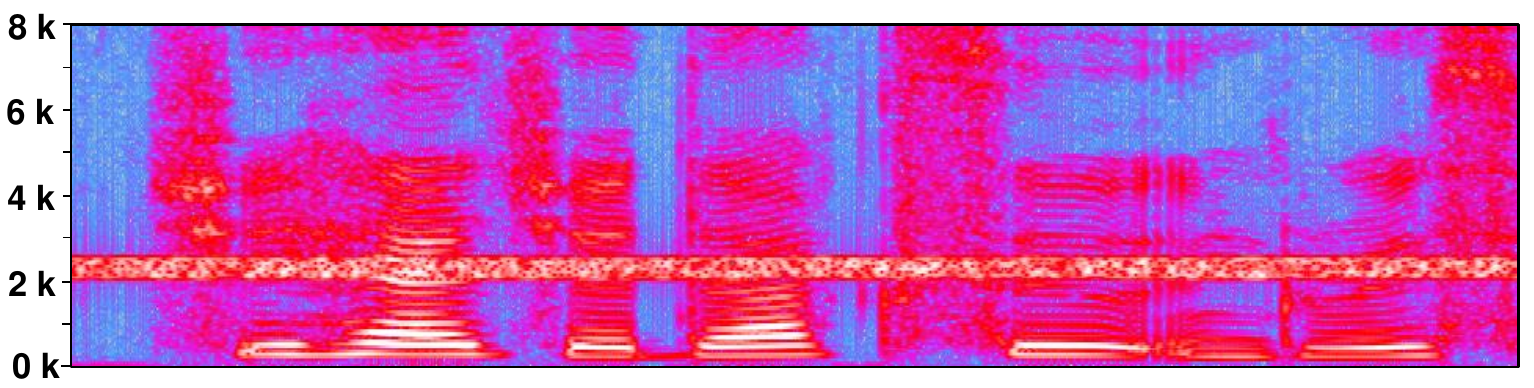}
\label{fig:cnn_filt}
\end{subfigure} \hspace{0.0\textwidth \vspace{0em}}

\begin{subfigure}{0.24\textwidth}
\includegraphics[scale=0.310,trim={0cm 0cm 0cm 0cm},clip]{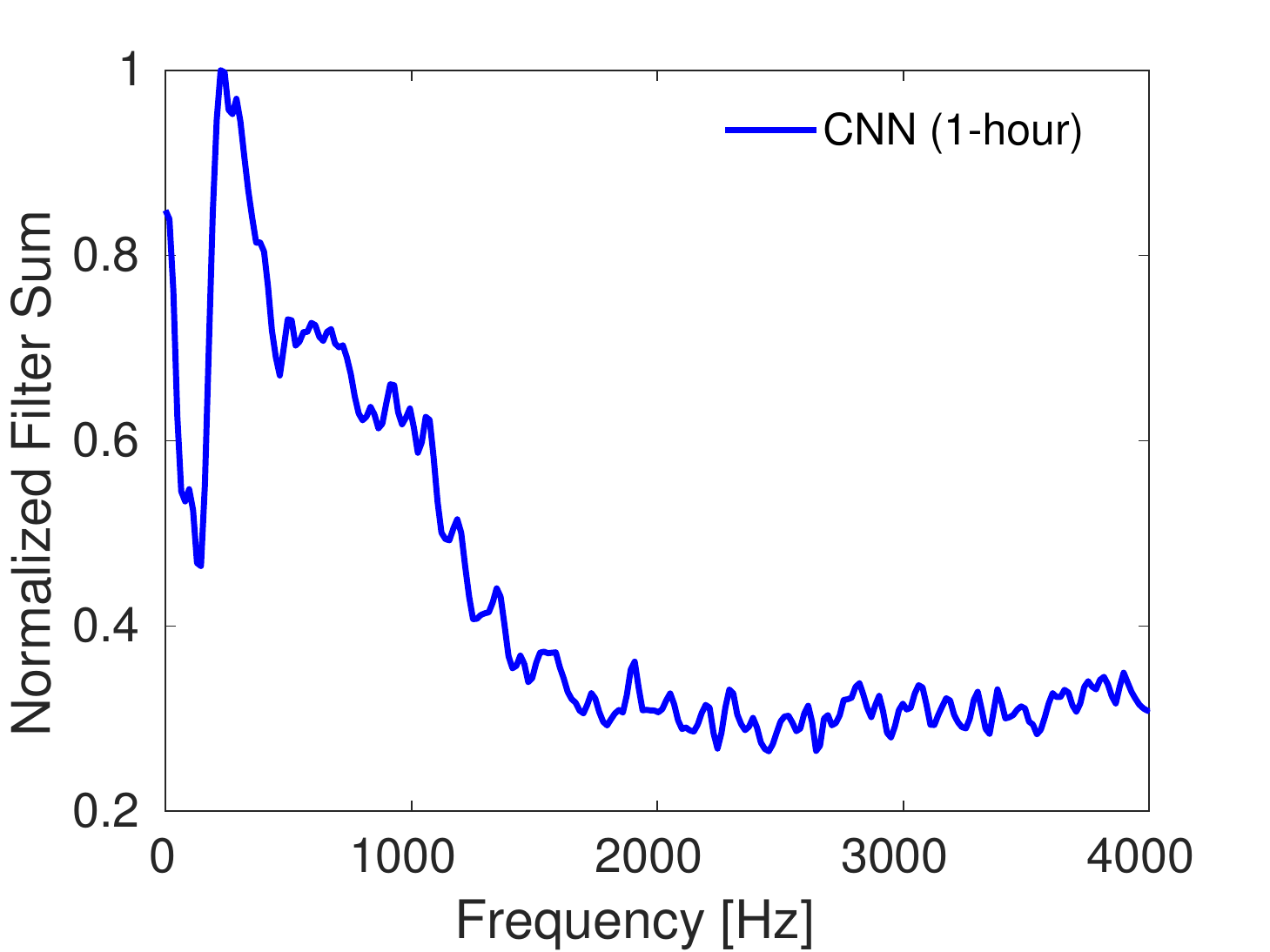}
\label{fig:cnn_filt}
\end{subfigure} \hspace{0.0\textwidth}
\begin{subfigure}{0.24\textwidth}
\includegraphics[scale=0.310]{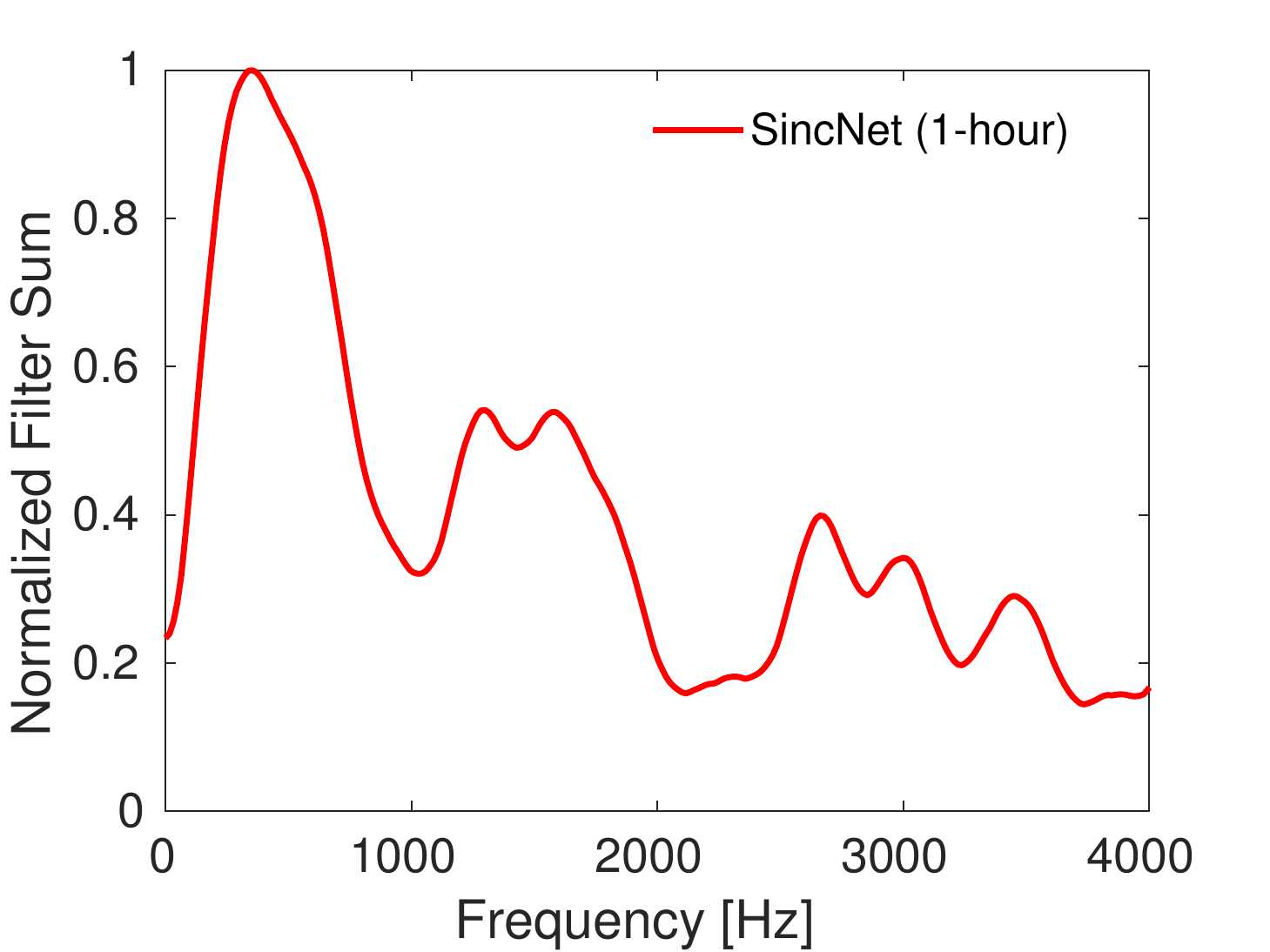}
\label{fig:sinc_filt}
\end{subfigure}

\begin{subfigure}{0.24\textwidth}
\includegraphics[scale=0.310,trim={0cm 0cm 0cm 0cm},clip]{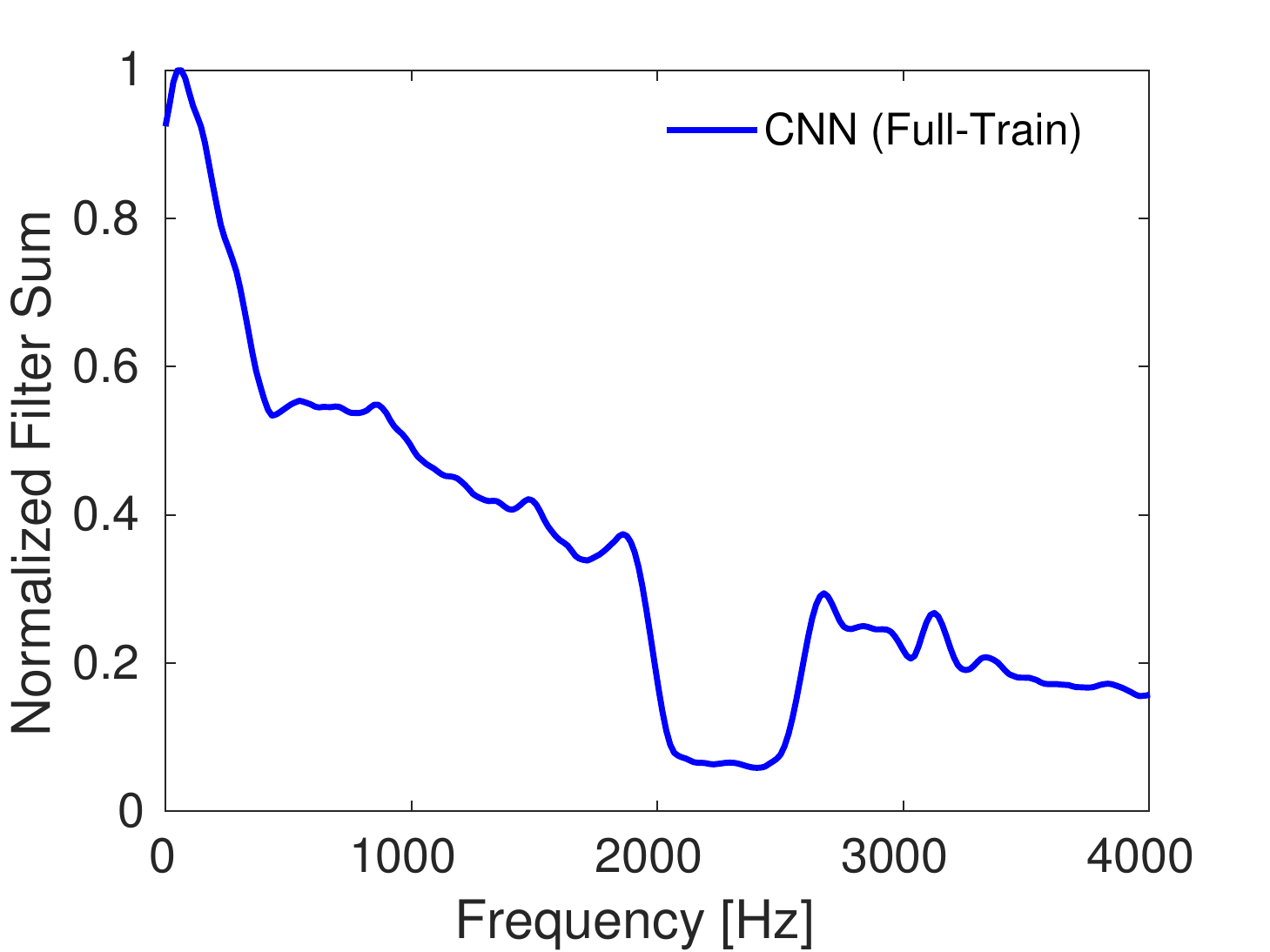}
\label{fig:cnn_filt}
\end{subfigure} \hspace{0.0\textwidth}
\begin{subfigure}{0.24\textwidth}
\includegraphics[scale=0.310]{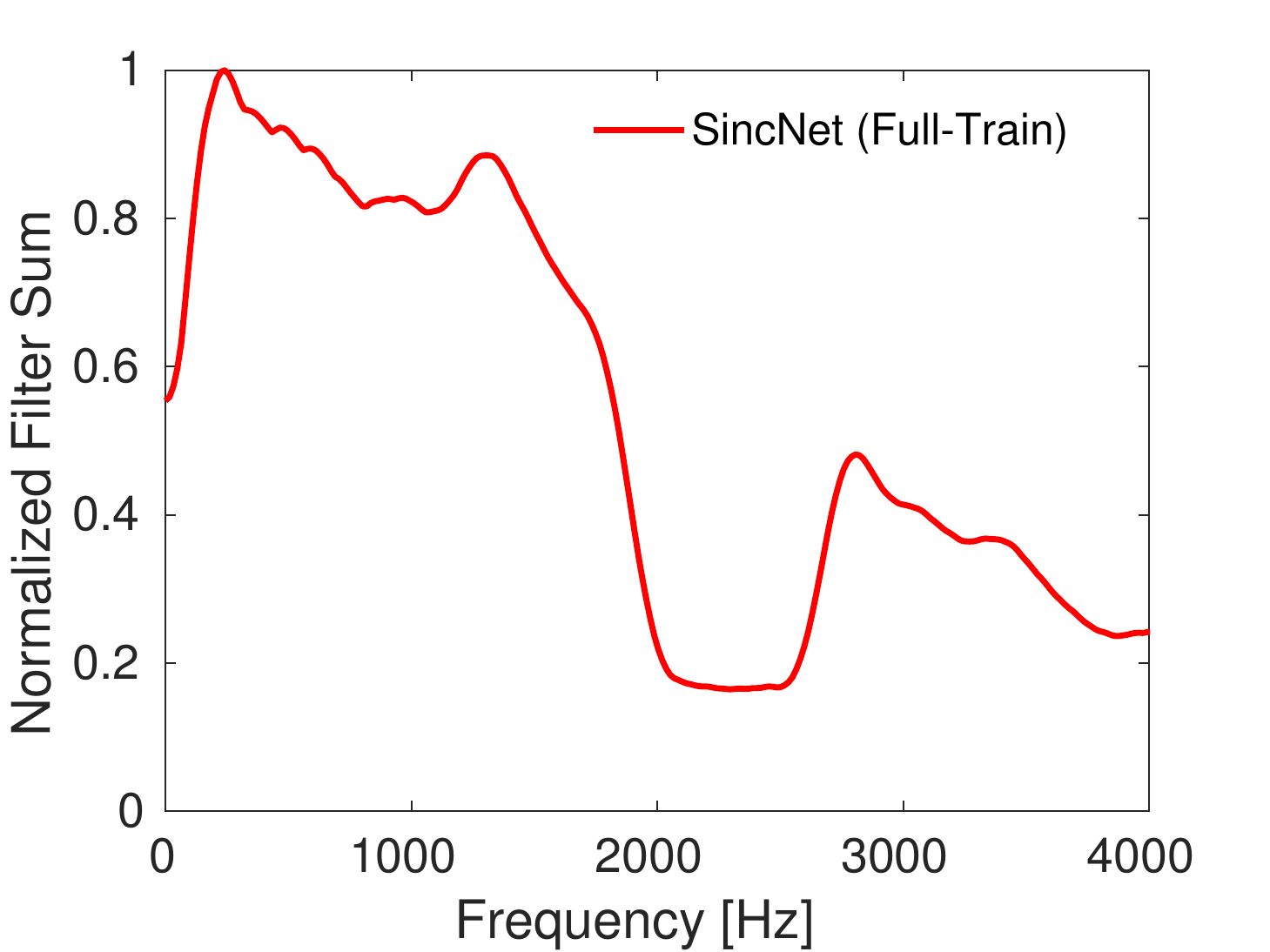}
\label{fig:sinc_filt}
\end{subfigure}

\caption{Cumulative frequency responses obtained on a noisy speech recognition task. As shown in the spectrogram, noise has been artificially added into the band 2.0-2.5 kHz. Both the CNN and SincNet learn to avoid the noisy band, but SincNet learns it much faster.}
\label{fig:asr_cum}
\end{figure}



\subsection{Speech Recognition}
To validate the effectiveness of our model, Fig. \ref{fig:asr_cum} shows the cumulative frequency response obtained on a noisy speech recognition task. In this experiment, reported here as a showcase, we have artificially corrupted TIMIT with a significant quantity of noise in the band between 2.0 and 2.5 kHz (see the spectrogram) and we have analyzed how fast a CNN and the SincNet architectures learn to avoid such a useless band. The second row of sub-figures compares the CNN and the SincNet at an early training stage (i.e., after processing only one hour of speech), while the last row shows the cumulative frequency responses at the end of training. From the figures emerges that both CNN and SincNet have correctly learned to avoid the corrupted band at end of training, as highlighted by the visible holes between 2.0 and 2.5 kHz in the cumulative frequency responses. SincNet, however, learns to avoid such a noisy band much earlier. In the second row of sub-figures, in fact, SincNet shows a valley in the cumulative spectrum even after processing only one hour of speech, while CNN will learn it only at a later training stage. At the end of training, the cumulative frequency responses of SincNet and CNN look rather different. SincNet, for instance, seems to exploit a larger bandwidth. Different from the CNN, in fact, it employs with several filters also the part of the spectrum between 0.5-2.0 kHz.

\begin{table}[h]
\caption{ASR performance obtained on the TIMIT and DIRHA.}
\centering

\begin{tabular}{llll}  
\toprule
& TIMIT &  DIRHA \\ 
& PER(\%)  &  WER(\%) \\ 
\midrule
CNN-FBANK       &   18.3      &  40.1    \\ 
CNN-Raw      &   18.1     &  40.1      \\ 
SincNet-Raw      &   \textbf{17.2}        &  \textbf{37.2}      \\ 
\bottomrule
\end{tabular}
\label{tab:asr}
\end{table}

Tab. \ref{tab:asr} reports the speech recognition performance obtained by CNN and SincNet using the TIMIT and the DIRHA dataset \cite{dirha_asru}. To ensure a more accurate comparison between the architectures, five experiments varying the initialization seeds were conducted for each 
model and corpus. Table \ref{tab:asr} thus reports the average speech recognition performance. Standard deviations, not reported here, range between $0.15$ and $0.2$ for all the experiments.

For all the datasets, SincNet outperforms both a CNN trained on standard FBANK coefficients and CNN trained or the raw waveform. The latter result confirms the effectiveness of SincNet not only in close-talking scenarios but also in noisy conditions characterized by the presence of both noise and reverberation. As shown in Fig.\ref{fig:asr_cum},  SincNet is able to effectively tune its filter-bank front-end to better address the characteristics of the noise. It is worth noting that the relative performance gain obtained in this challenging scenario is slightly higher than that obtained in standard close-talking conditions (6\% against 4\%).

\section{Conclusions and Future Work}
\label{sec:conc}
This paper proposed SincNet, a neural architecture for directly processing waveform audio. Our model, inspired by the way filtering is conducted in digital signal processing, imposes constraints on the filter shapes through efficient parameterization. SincNet has been extensively evaluated on challenging speaker and speech recognition task, consistently showing some benefits on network convergence, performance, and computational efficiency. 
Moreover, analysis of the SincNet filters revealed that the learned filter-bank is tuned to the specific task addressed by the neural network.

Inspired by the promising results obtained in this paper, we will explore the use of SincNet for supervised and unsupervised speaker/environmental adaptation. 
We believe that the proposed approach defines a general paradigm to process time-series and can be applied in numerous other fields. 
Our future effort will be thus devoted to extending to other tasks, such as emotion recognition, speech separation, and music processing. 


\vfill\pagebreak

\bibliographystyle{IEEEbib}
\bibliography{ref}

\end{document}